\documentclass[12pt]{iopart}

\begin{document}

\jl{31}

\title{Exploring a rheonomic system}

\author{Antonio S de Castro}

\address{UNESP - Campus de Guaratinguet\'a - Caixa Postal 205 - 12500000
Guaratinguet\'a - SP - Brasil}

\begin{abstract}
A simple and illustrative rheonomic system is explored in the Lagrangian
formalism. The difference between Jacobi's integral and energy is
highlighted. A sharp contrast with remarks found in the literature is
pointed out. The non-conservative system possess a Lagrangian not
explicitly dependent on time and consequently there is a Jacobi's
integral. The Lagrange undetermined multiplier method is used as a
complement to obtain a few interesting conclusions.
\end{abstract}

\pacno{03.20.+i}

\submitted

\maketitle

Constraints are restrictions that limit the motion of the particles of a
system. The forces necessary to constrain the motion are said to be
forces of constraint. The constraints expressible as algebraic equations
relating the coordinates of the particles and the time variable are called
holonomic, if not 
they are called nonholonomic. Furthermore, in each type of constraints, 
holonomic or nonholonomic, the time variable could appear explicitly. If the
time variable does not appear explicitly in the relations of constraint
they are further classified as scleronomic, otherwise they are said to be
rheonomic. 

Holonomic constraints, and in fact a very
restrict class of nonholonomic constraints (those expressible as
first-order differential forms relating the coordinates and the time
variable), are amenable to straightforward general treatment in analytical
mechanics. These sorts of constraints allow us to describe the motion
without paying any explicit reference to the forces of constraint. In
addition, holonomic constraints can be used to reduce the number of
coordinates required to  the complete description of the motion, although
this is not always desirable. 

Simple systems subject to rheonomic constraints are not widespread in the
textbooks on analytical mechanics. Nevertheless, there is a traditional
system which is very simple, indeed. It consists of a bead of mass $m$
sliding along a frictionless straight horizontal wire constrained to
rotate with 
constant angular velocity $\omega $ about a vertical axis \cite{o}\cite{k}%
\cite{l}\cite{g}\cite{c}. This simple system presents a wealth of physics
not fully explored in the literature. The main purpose of this paper is to
make an effort for filling this gap, motivated by the strong pedagogical
appeal of this illustrative system. Furthermore, this paper takes the
opportunity of doing criticisms on the remarks in Griffths's textbook \cite
{gr} concerning general systems containing rheonomic holonomic systems: ``%
\textit{...the rheonomic constraints must be used to reduce the number of
generalised coordinates and so the configuration of the system must
necessarily depend explicitly on time as well as the n generalized
coordinates. In this case a time dependence thus enters explicitly into the
Lagrangian. It may therefore also be concluded that systems which contain a
rheonomic constraint possess neither an energy integral nor a Jacobi
integral.}''

First, one can note that the motion of the bead is caused by a force of
constraint perpendicular to the wire, whereas the actual displacement of the
bead is in an oblique direction and its virtual displacement satisfying the
constraint is in a parallel direction. Therefore, the force of constraint
does  actual work but not  virtual work. The vanishing of the virtual work
characterizes the constraint as ideal.

Since the motion of the bead takes place on the horizontal plane one can
eliminate the dependence on the vertical coordinate and consider only the
coordinates on the plane of the motion. The coordinates are suitably chosen
with $r$ being the distance to the rotation axis and $\theta $ the angular
position relative to an arbitrary axis on the plane of the motion. The
Lagrangian of the system is nothing but the kinetic energy of the bead:

\begin{eqnarray}
L=\frac{1}{2}m\left( \dot{r}^{2}+r^{2}\dot{\theta}^{2}\right)  \label{l1}
\end{eqnarray}
The constraint on the motion of the bead is expressed by

\begin{eqnarray}
\phi \left( \dot{\theta} \right) =\dot{\theta} -\omega =0
\end{eqnarray}
This relation can be immediately integrated, yielding

\begin{eqnarray}
\Phi \left( \theta ,t\right) =\theta -\omega t+\theta _{0}=0  \label{v}
\end{eqnarray}
where $\theta _{0}$ is a constant. This form of the condition of constraint
allow us to classify it as a holonomic and rheonomic constraint. Now one
can use 
this condition of constraint to eliminate the coordinate $\theta $ in the
Lagrangian, so that one is left with $r$ as generalized coordinate:

\begin{eqnarray}
L=\frac{1}{2}m\left( \dot{r}^{2}+\omega ^{2}r^{2}\right) \label{l2}
\end{eqnarray}
At this point the author dares to utter the first criticism on Griffths's
conclusions. The configuration of the system is just given by the coordinate 
$r$, which clearly depends on the time variable, but neither the kinetic energy
nor the Lagrangian are explicitly time-dependent. In general rheonomic
constraints give rise to explicitly 
time dependent terms in the Lagrangian. There are two of these terms, one of
them is linear in the generalized velocities and the other one is
velocity-independent. Due to these terms it may tempting to conclude that
the Lagrangian for a rheonomic system is always explicitly time dependent,
but it is very dangerous because it may be certain cancellations. For the
particular system 
approached in this paper and with the particular choice of generalized
coordinates, the variables combine in such a way that the linear term
vanishes whereas the independent term does not involve the time explicitly.
That is the reason why the Lagrangian has no explicit time dependence.

Using the Lagrangian given by (\ref{l2}) Lagrange's equation governing the
motion of the bead

\begin{eqnarray}
\frac{d}{dt}\left (\frac{\partial L}{\partial \dot{r}}\right )- \frac{%
\partial L}{\partial r}=0
\end{eqnarray}
takes on the form

\begin{eqnarray}
m\ddot{r}-m\omega ^{2}r=0  \label{m}
\end{eqnarray}
The energy function $h$, generally given by

\begin{eqnarray}
h=\sum_{i}\dot{q}_{i}\frac{\partial L}{\partial \dot{q}_{i}}-L
\end{eqnarray}
obeys the relation

\begin{eqnarray}
\frac{dh}{dt}=-\frac{\partial L}{\partial t}
\end{eqnarray}
and for the present system it is given by

\begin{eqnarray}
h=\dot{r}\frac{\partial L}{\partial \dot{r}}-L
\end{eqnarray}
Since the Lagrangian is not an explicit function of time

\begin{eqnarray}
h=\frac{1}{2}m\left( \dot{r}^{2}-\omega ^{2}r^{2}\right) 
\end{eqnarray}
turns out to be Jacobi's integral, a constant of the motion. Now arises the
second criticism on Griffiths's comments: although this system contains a
rheonomic constraint it in fact possess a Jacobi's integral. The necessary
and sufficient condition for the existence of Jacobi's integral is that the
Lagrangian does not depend explicitly on time. It is seen that here Jacobi's
integral is not the energy of the system. The only difference between them
is due to the velocity-independent term in the Lagrangian. The energy of the
system is only kinetic energy and has a time derivative given by

\begin{eqnarray}
\frac{dE}{dt}=m\dot{r}\left( \ddot{r}+\omega ^{2}r\right)
\end{eqnarray}
The insertion of the equation of motion (\ref{m}) into the last relation
leads to

\begin{eqnarray}
\frac{dE}{dt}=\frac{d}{dt}\left( m\omega ^{2}r^{2}\right)= 2m\omega^2 r\dot{r%
}  \label{e}
\end{eqnarray}
which implies that the energy is not a constant of the motion. As we have
already seen, the energy can not be a constant of the motion due to the
nonvanishing of the actual work of the force of constraint. 

It should be obvious that the energy function $h$ and the energy $E$ are
distinctly different functions, subject to distinct conservation laws, but
there are special circumstances for which they are identical. This happens
if the constraints are scleronomic and the potential energy is
velocity-independent. If, further, the potential energy does not depend
explicitly on time $E$ becomes the energy integral and $h$ comes to be
Jacobi's integral. In addition to 
these comments is appropriated to keep in mind that the energy function
$h$ must not be confused with the Hamiltonian $H$, even though they are
expressed by similar mathematical structures and their conservation laws
rest upon the very same condition (not depend explicitly on time).
The difference between $h$ and $H$ is subtler than that one between $h$
and $E$, they are functions of different independent variables. As a
matter of fact, in some  cases it may not be possible to obtain one
of them from the knowledge from the other.

Usually one must use the Lagrange undetermined multiplier method to obtain
the force of constraint. In this method the coordinates $r$ and $\theta$ are
not treated as independent coordinates, therefore one has to use the
Lagrangian given by (\ref{l1}) instead of that one given by (\ref{l2}). Now
Lagrange's equations incorporate the condition of constraint

\begin{eqnarray}
\frac{d}{dt}\left (\frac{\partial L}{\partial \dot{r}}\right )- \frac{%
\partial L}{\partial r}=\lambda \frac{\partial \Phi }{\partial r}
\end{eqnarray}

\begin{eqnarray}
\frac{d}{dt}\left (\frac{\partial L}{\partial\dot{\theta}}\right )- \frac{%
\partial L}{\partial \theta }=\lambda \frac{\partial \Phi }{\partial \theta }
\end{eqnarray}
where $\lambda $ is the Lagrange undetermined multiplier. The generalized
forces of constraint are to be identified as $\lambda \partial \Phi
/\partial r$ and $\lambda \partial \Phi /\partial \theta$. The condition of
constraint (\ref{v}) implies that only the torque of constraint $\tau$ is
nonvanishing. These Lagrange's equations yield

\begin{eqnarray}
m\ddot{r}-m\omega ^{2}r=0
\end{eqnarray}

\begin{eqnarray}
\tau =\frac{dp_{\theta }}{dt}=mr\left( 2\dot{r}\dot{\theta}+r\ddot{\theta}%
\right)  \label{t1}
\end{eqnarray}
where

\begin{eqnarray}
p_{\theta }=\frac{\partial L}{\partial \dot{\theta}}=mr^{2}\dot{\theta}
\label{p}
\end{eqnarray}
happens to be the angular momentum. Combining (\ref{t1}) and (\ref{p}) with (%
\ref{v}) one gets

\begin{eqnarray}
\tau=2m\omega r\dot{r}  \label{t2}
\end{eqnarray}

\[
p_{\theta }=m\omega r^{2} 
\]
That the angular momentum is not a constant of the motion comes from the
fact that the force of constraint is not central. The constraint force can
now be obtained from (\ref{t2}) reckonizing that it acts on the bead
directed normal to the wire. It is an easy matter to check that

\begin{eqnarray}
F_{\theta }=2m\omega\dot{r}
\end{eqnarray}

In conclusion, this paper shows that the system considered is of great value
for beginning students of analytical mechanics. In addition, it is very
useful to remove some misunderstandings found in the literature. It should
be emphasized that is the rheonomic nature of the constraint and the
particular choice of generalized coordinates that make the energy to be
different from Jacobi's integral. Jacobi's integral is here the first
integral of the motion instead of the energy. The Lagrange undetermined
multiplier method has been used for obtaining the force of constraint in a
natural way. Nonetheless, the force of constraint can also be obtained from (%
\ref{e}) by invoking the principle of work and energy. Only the Lagrangian
formalism has been considered in this paper but this simple system can also
be easily approached by other formalisms of the analytical mechanics. This
task is left to the readers. \newline
\newline

\end{document}